\documentclass[pra,aps,twocolumn,superscriptaddress,showpacs]{revtex4-1}
\usepackage[T1]{fontenc}
\usepackage[latin9]{inputenc}
\usepackage{geometry}
\usepackage{amssymb}
\usepackage{graphicx}
\usepackage{esint}
\usepackage{array}
\usepackage{multirow}

\begin{document}

\title{Measurement-based tailoring of Anderson localization of
partially coherent light}

\author{Ji\v{r}\'{i} Svozil\'{i}k}
\email{jiri.svozilik@icfo.es} \affiliation{ICFO-Institut de
Ciencies Fotoniques, Mediterranean Technology Park, 08860,
Castelldefels, Barcelona, Spain} \affiliation{Palack\'{y}
University, RCPTM, Joint Laboratory of Optics, 17. listopadu 12,
771 46 Olomouc, Czech Republic}

\author{Jan Pe\v{r}ina Jr.}
\affiliation{Palack\'{y} University, RCPTM, Joint Laboratory of
Optics, 17. listopadu 12, 771 46 Olomouc, Czech Republic}

\author{Juan P. Torres}
 \affiliation{ICFO-Institut de
Ciencies Fotoniques, Mediterranean Technology Park, 08860,
Castelldefels, Barcelona, Spain} \affiliation{Department of Signal
Theory and Communications, Universitat Politecnica Catalunya,
Campus Nord D3, 08034 Barcelona, Spain}

\begin{abstract}
We put forward an experimental configuration to observe 
transverse Anderson localization of partially coherent light beams 
with a tunable degree of first-order coherence. The scheme 
makes use of entangled photons propagating in disordered 
waveguide arrays, and is based on the unique relationship 
between the degree of entanglement of a pair of photons and 
the coherence properties of the individual photons constituting 
the pair. The scheme can be readily implemented with current 
waveguide-on-a-chip technology, and surprisingly, the tunability 
of the coherence properties of the individual photons is done at 
the measurement stage, without resorting to changes of the 
light source itself.
\end{abstract}

\pacs{42.50.Dv, 42.25.Dd,72.15.Rn}

%42.50.Dv   Quantum state engineering and measurements
%42.25.Dd   Wave propagation in random media
%72.15.Rn   Localization effects (Anderson or weak localization)

\maketitle

\section{Introduction}

%Anderson
More than 50 years ago, P. W. Anderson described in a seminal
paper \cite{Anderson1958Absence} how diffusion in the process
of electron transport in a disordered ({\em random})
semiconductor lattice can be arrested, leading to the localization
of the wavefunction in a small region of space, the so-called {\em
Anderson localization}. This unique phenomenon has been
observed in a myriad of physical systems
\cite{Anderson2009PhysToday}, including electron gas
\cite{Cutler1969Observation}, matter waves (atoms)
\cite{Chabe2008Experimental,Billy2008Direct,Roati2008Anderson},
 and acoustic waves \cite{Hu2008Localization}. The observation
of transverse localization of light in a photonic system was
predicted by De Raedt et al. \cite{Raedt1989Transverse},
considering the similarities existing between the Schr\"{o}dinger
equation and Maxwell equations. This led to the observation of
Anderson localization in photonic systems
\cite{Wiersma1997Localization,Schwartz2007Transport,Schreiber2011Decoherence,Crespi2013Anderson,Lahini2008Anderson}
in various scenarios.

%Quantum Random Walk
The underlying physical principles that lead to Anderson
localization are also responsible for changes on the spreading of
the wavefunction in a quantum random walk, characterized by a
quadratic dependence of the size (variance) of the wavefunction
with propagation distance when no disorder is present
\cite{Aharonov1993Quantum,Svozilik2012Implementation}. The
consequences of introducing static disorder in a quantum random
walk (leading to Anderson localization) have been studied, for
example, for one-dimensional
\cite{Yin2008Quantum,PerinaJr2009b,PerinaJr2009c} and
two-dimensional \cite{Inui2004Localization} systems. In a sense,
generalizations of quantum protocols such as the Shor's
factorization algorithm \cite{Shor1994Algorithms} and the
Groover's searching algorithm \cite{Grover1996Afast} can also 
be analyzed in similar terms, since they can be viewed as 
quantum random walks.

%Partially coherent light
In most cases, the input state in a quantum random walk is
considered to be fully coherent. Since Anderson localization is a
consequence of interference effects, one can dare thinking that 
an initial coherent state is thus necessary to observe Anderson
localization. However, \v{C}apeta et al. 
\cite{Capeta2011Anderson} have shown that even a partially 
coherent input light beam can lead to Anderson localization in a 
disordered waveguide array (WGA). Partially coherent beams 
can be described as a superposition of orthogonal coherent 
modes, where the modal coefficients are random variables that 
are uncorrelated with one another 
\cite{mandel_book,segev2001}. Therefore, according to
\cite{Capeta2011Anderson}, since spreading of each mode, 
being a coherent mode, can be arrested in a random medium 
with static disorder, the whole partially coherent beam should 
also suffer localization in a similar way to a fully coherent beam.

%Our goal
Here we propose an experimental scheme which could lead to 
the observation of Anderson localization of partially coherent 
beams with a tunable degree of first-order coherence. The 
approach is based on two basic ingredients. On the one hand, a 
single-photon in a pure quantum state (von Neumann entropy 
$E=0$) is arguably the most simple example of a photonic state 
which shows first-order coherence \cite{glauber1966}. Mixed 
single-photon quantum states do not show first-order 
coherence. On the other hand, the degree of entanglement of a 
pure two-photon state (photons A and B) is directly related to 
the purity of the quantum state of photon A (B), which 
results from tracing out all degrees of freedom corresponding to 
photon B (A). The von Neumann entropy of the quantum state 
that describes photon A (B) could be used as a measure of the 
degree of entanglement of the paired photons.

Consequently, the manipulation of the degree of entanglement of
the two-photon state can effectively tailor the first-order
coherence of the signal (idler) photon \cite{victor2009},
generating a one-photon quantum state which is mixed, and thus
partially coherent. Anderson localization ({\em co-localization})
of entangled photon fields in disordered waveguides has been
presented in
\cite{Lahani2010Quantum,Abouraddy2012Anderson,DiGiuseppe2013Einstein}.
However, in that case the goal was to look for Anderson
localization of the two photons that form the entangled pair,
while here entanglement is a tool to tailor the degree of
coherence of one of the subsystems (photon A or photon B)
which form the entangled pair.

The paper is organized as follows. In Section II the experimental
scheme and the main theoretical tools used in the analysis are
presented and discussed. Main results are presented in Section
III. The single-photon coherence measures used throughout the
text are defined in the Appendix.

\section{The proposed experimental scheme}
In general, the quantum description of a pure entangled two-photon
state (photons A and B) is written
\begin{equation}
\label{two-photon-state1}
 |\Psi\rangle=\int d p\int  dq\,   \Psi(p,q) \hat{a}_A^{\dagger}(p) 
 \hat{a}_B^{\dagger}(q) |0 \rangle
\end{equation}
where $p$  and $q$ represent the transverse wavevectors of 
photons $A$ and $B$, respectively, $ \hat{a}_A^{\dagger}(p)$ 
and $ \hat{a}_A^{\dagger}(q)$  are creation operators of 
photons in  modes $A$ and $B$, and $\Psi(p,q)$ is the mode 
function that describes the properties of the biphoton 
\cite{progress2011}. For monochromatic fields, the 
positive-frequency electric-field operators are expressed as
\begin{eqnarray}
&  & \hat{E}_A^{(+)}(x) \sim \int dp\, \hat{a}_A(p) \exp \left( i p x\right),
 \\
&  & \hat{E}_B^{(+)}(y) \sim \int dq\, \hat{a}_B(q) \exp \left( i
q y\right) .
\end{eqnarray}
We note that temporal dependence of the electric-field operators
has been omitted for the sake of simplicity. Defining $\Psi(x,y) =
\int dp\,\int dq\, \Psi(p,q) \exp(-ipx-iqy)$, the normalized pure
entangled two-photon state given by Eq. (\ref{two-photon-state1})
can be written as
\begin{equation}%Eq:1
\label{state1} |\Psi\rangle=\int dx\int dy\,\Psi\left(x,
y\right)|x\rangle_A |y\rangle_B \label{two-photon-state2},
\end{equation}
where we have defined $|x\rangle_A \equiv \hat{E}_{A}^{(-)}
\left(x\right)|0\rangle_A $ and $|y\rangle_B \equiv
\hat{E}_{B}^{(-)}\left(y\right)|0\rangle_B $. Notice that the
two-photon amplitude $\Psi(x,y)$ corresponds to the second-order
correlation function $ \Psi(x,y) = {}_A\langle 0| {}_B\langle 0|
\hat{E}_B^{(+)}(y) \hat{E}_A^{(+)}(x) \rangle$.

The two-photon amplitude $\Psi$ can be described by a Schmidt
decomposition of the form
\begin{equation}
\label{schmidt} \Psi(x,y)=\sum_{j=1}^N \sqrt{\lambda_j}
f_j\left(x\right) g_j \left(y \right);
\label{Eq:2}
\end{equation}
$\lambda_j$ are the Schmidt eigenvalues and $\{f_j\}$ and
$\{g_j\}$ are the Schmidt modes corresponding to photons A and B.
For the sake of simplicity, the two-photon amplitude $\Psi(x,y)$
is approximated by the Gaussian function
\begin{equation}
\label{gaussian_mode_function}
\Psi(x,y)\sim\exp\left[-\alpha\left(x+y\right)^{2}-\beta\left(x-y\right)^{2}\right].
\label{Eq:3}
\end{equation}

\begin{figure}[t]
\centering
\includegraphics[width=8.70cm]{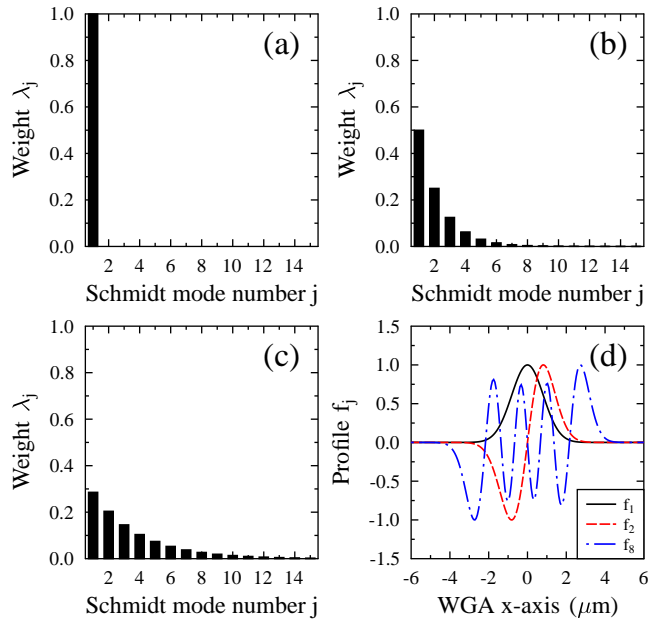}
 \caption{(Color online) Schmidt mode decomposition. Here we
 show the first $15$ Schmidt modes of the Schmidt
 decomposition for three cases: a) a separable state with
$\gamma_0=0.5$, (b) an entangled state with 
$\gamma_0=1.5$, and (c) an entangled state with 
$\gamma_0=3$. The shape of some selected
Schmidt modes (see the legend) are plotted in (d) for
$\gamma_0=1.5$. In all cases $\sigma_0$ = 1 $\mu m$.}
\label{Fig:1}
\end{figure}

\noindent In this case, the Schmidt modes correspond to Hermite
functions of order $j$ \cite{chan_eberly,PerinaJr2008}. Some
representative cases are shown in Fig.\ref{Fig:1}(d).

The parameters characterizing the spatial correlations between
photons $A$ and $B$, $\alpha$ and $\beta$, can be expressed 
using more suitable parameters that describe characteristics of 
photon A: its rms beam width ($\sigma_{0}$) and the beam 
width-spatial bandwidth product ($\gamma_0$), here denoted 
as incoherence, 

\begin{eqnarray}
\alpha & = & \frac{1}{4\sigma_{0}^{2}}\left(2\gamma_0^{2}\pm\gamma_0\sqrt{4\gamma_0^{2}-1}\right), \label{Eq:4}\\
\beta  & = &
\frac{1}{4\sigma_{0}^{2}}\left(2\gamma_0^{2}\mp\gamma_0\sqrt{4\gamma_0^{2}-1}\right).
\label{Eq:5}
\end{eqnarray}
The derivation of Eq.~(\ref{Eq:4}) and Eq.~(\ref{Eq:5}) is
included in the Appendix. In general, $\gamma_0 \ge 0.5$ and is
related to the Schmidt number, $K=2\gamma_0$, which is a 
measure of the size of the mode distribution involved in Eq.
(\ref{schmidt}), i.e., $K= (\sum_{j=1}^{N}
\lambda_j)^2/\sum_{j=1}^{N} \lambda_j^2$. For 
$\alpha=\beta$,  there is not entanglement between photons A 
and B, the Schmidt decomposition contains a single mode [see 
Fig.~\ref{Fig:1}(a)], and $\gamma_0$ attains its minimum 
value, i.e., $\gamma_0=0.5$. This case yields a pure and 
first-order coherent photon. In all other cases, the spectrum of 
the Schmidt decomposition contains several modes. 
Fig.\ref{Fig:1}(b) shows the weights of the first $15$ Schmidt 
modes (eigenvalues $\lambda_j$) for $\gamma_0=1.5$ and 
Fig.\ref{Fig:1}(c) shows them for $\gamma_0=3$.

The key point of our scheme is the presence of a detection 
scheme that projects the photon $B$ into a restricted set of 
modes before detection, or in particular the projection 
into a single Schmidt mode $g_j$. In this way, the number of 
modes that describe the quantum state of photon A after 
detection of photon $B$ would be correspondingly reduced. 
Importantly, the first-order coherence of photon $A$ depends on 
the number of modes onto which the photon $B$ is projected. 
The projection of photon $B$ into a specific single mode 
effectively 
renders photon $A$ into a first-order coherent
photon. In contrast, detection of photon $B$ into an
increasing number of modes results in a partially coherent
signal photon with a decreasing degree of coherence. Therefore,
this can thus be appropriately called tailoring of the first-order
coherence by heralding detection.

By tailoring the first-order coherence of a single photon, we also
tailor the characteristics of the Anderson localization. The
projection and detection of photon $B$ into a finite number $M$
of modes is represented by the quantum operator
$\hat{Y}_B=\sum_{j=1}^{M}|g_j\rangle_B\langle g_j|_B$ with
$|g_j\rangle=\int dy g_j(y)|y\rangle_B$. After detection, the {\em
truncated} quantum state of photon A reads as
\begin{equation}
\label{signal_coherence} \hat{\rho}_A={\rm
Tr}_B\left[|\Psi\rangle\langle\Psi|\hat{Y}_B\right]=\sum_{j=1}^{min\left(N,M\right)}
 \lambda_j
|f_j\rangle_A\langle f_j|_A
\label{Eq:6}
\end{equation}
corresponding to an incoherent superposition of $min\left(N,M\right)$ modes with
weights $\lambda_j$.

\begin{figure}
\centering\includegraphics[width=8cm]{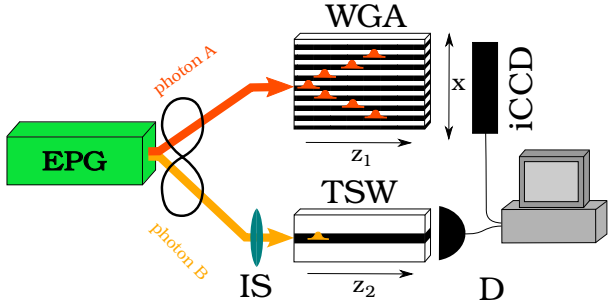} \caption{(Color
online) Sketch of the experimental configuration proposed to
observe Anderson localization of partially coherent photons in a
disordered waveguide array (WGA). The three-slab waveguide
(TSW) allows propagation of different numbers of guided modes
depending on its core size. EPG: The Entangled-photons
generator is the source of photon pairs; iCCD: intensified CCD;
D: single-photon detector; IS: Imaging System.}
\label{Fig:2}
\end{figure}

A sketch of the experimental configuration considered is shown in
Fig.~\ref{Fig:2}. A pair of entangled photons (A and B) is
generated.  Photon A is injected into a one-dimensional
waveguide array (WGA) with refractive index profile
$n_{A}(x)$. The waveguide array contains $101$ layers of
semiconductor material $Al_{x}Ga_{1-x}As$ with the index of
reflection taken from \cite{Gehrsitz2000The}. The whole
structure is created by alternating two different layers:
$Al_{0.3}Ga_{0.7}As$ and $Al_{0.8}Ga_{0.2}As$ of the
same thickness 0.6 $\mu m$. The disorder is induced by
randomizing the index of refraction of each layer, etc.:
$n_{A}(x)=n^{0}_{A}(x)+\Delta n_A(x)$. The probability
distribution of the random disturbances $\Delta n_A(x)$ is
described by a Gaussian function characterized by its typical
standard deviation $\delta$.

\begin{figure}[t]
\centering
\includegraphics[width=8.70cm]{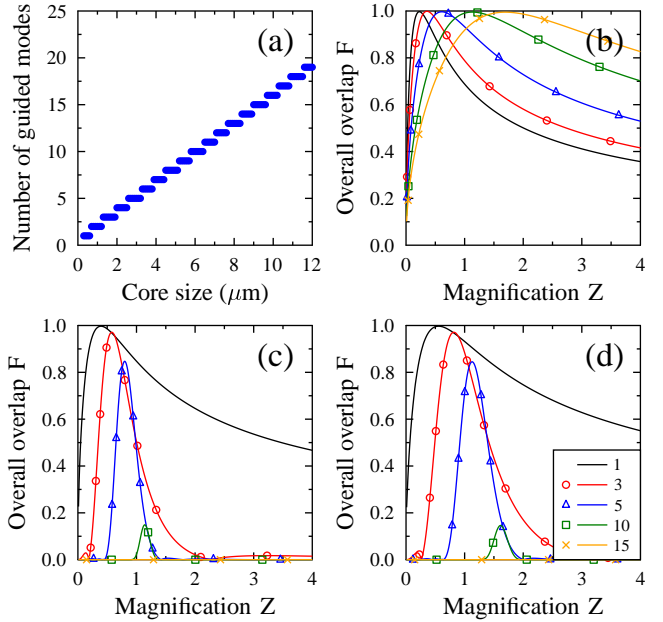}
\caption{(Color online) (a) Number of guided modes supported
by the three-slab waveguide (TSW) as a function of the core size
of the waveguide. (b)-(d) show the overall spatial
overlap factor between Schmidt modes $\{g_j\}$ and guided
modes of the three-slab  waveguide $\{v_j\}$, as given by the
product $F=\prod_j|d_{jj}|$,  as a function of the magnification
factor Z  of  the imaging system.  Three different cases, with
different values  of $\gamma_0$, are considered: (b)
$\gamma_0=0.5$, (c)  $\gamma_0=1.5$, and (d)
$\gamma_0=3$. In all cases $\sigma_0$ = 1 $\mu m$. The
five curves in each plot correspond to five different three-slab
waveguides supporting various amounts of modes, as given by
the legend in (d).}
\label{Fig:3}
\end{figure}

On the other hand, the photon B can propagate in different
three-slab waveguides (TSWs) with refractive index profile
$n_{B}(y)$, and different sizes of the core of the waveguide. 
The material of the core is $Al_{0.3}Ga_{0.7}As$ and two 
surrounding layers are made of $Al_{0.8}Ga_{0.2}As$. The 
layers surrounding the core are considered to be infinite in their 
thickness.  The number of guided modes supported depends on 
the core size [see Fig.\ref{Fig:3}(a)], so the three-slab 
waveguide effectively selects a certain amount of modes of 
photon B, effectively tailoring the first-order coherence of photon 
A. A three-slab waveguide has been chosen for simplicity, and 
because of its suitability for integration on a chip altogether with 
the WGA.

The evolution of the spatial shape of photons A and B, in the
waveguide array and the three-slab waveguide, respectively, can
be conveniently described by means of the guided modes
supported by each waveguide, $\{u_{i}(x)\}$ for the WGA and
$\{v_{j}(y)\}$ for the TSW  \cite{Karbasi2013Modal}. The
guided modes are obtained as solutions of the Helmholtz
equations
\begin{eqnarray}
& & \Delta
u_i(x)+\left[n_{A}^{2}(x)k_{0}^{2}-\kappa_i^{2}\right]u_i(x)=0,
\label{Eq:7} \\
 & & \Delta v_j(y)+\left[n_{B}^{2}(y)k_{0}^{2}-\mu_j^{2}\right]v_j(y)=0
\label{Eq:8},
\end{eqnarray}
where $\kappa_i$ and $\mu_j$ are the corresponding
propagation constants. The index of refraction is considered to
be homogeneous along the direction of propagation (along the
$z$-axis in both waveguides). Equation (\ref{Eq:7}) has been
solved using the finite element method \cite{Jin2002Finite},
whereas Eq.~(\ref{Eq:8}) has been solved by the
semi-analytical method \cite{Snyder1983Optical}. The
polarization of photons A and B is transverse electric, i.e.
parallel to the surface boundary between layers, and their
wavelengths are $1550$ nm, far below the band gap of the
material. Therefore, absorption can be omitted in our model.
Moreover, the propagation distance $z_1$ of photon A
has been restricted to 0.5 mm in order to prevent reaching the
reflective boundaries of WGA.

The coupling of the input photons, characterized by the Schmidt
modes $f_n$ and $g_m$, to the corresponding waveguides,
characterized by modes $u_i$ and $v_j$, is expressed via the
coupling coefficients
\begin{eqnarray}
& & c_{ni}=\int dx f_n(x) u_{i}^{*}(x), \label{Eq:9} \\
& & d_{mj}=\int dy g_m (y) v_{j}^{*}(y) \label{Eq:10}.
\end{eqnarray}
Using coefficients $ c_{ni} $ and $ d_{mj} $ the quantum
state of
two photons after their propagation at distances $z_1$ and $z_2$
in the two waveguides is
\begin{eqnarray}
& & |\Psi\rangle=\sum_{n} \sqrt{\lambda_{n}} \sum_{ij} c_{ni}
d_{n_j} \nonumber \\
& & \times \exp \left( i\kappa_i z_1+i\mu_j z_2 \right)
|u_i\rangle_{A} |v_j\rangle_{B},
 \label{Eq:11}
\end{eqnarray}
where $|u_i\rangle_{A} \equiv \int dx u_i(x) |x\rangle_{A}$ and
$|v_j\rangle_{B} \equiv \int dy v_j(y) |y\rangle_{B}$. We can
write $z_1=z_2=z$ without losing generality.

Detection of photon B after projection via a three-slab waveguide
is represented by the operator
$\hat{Y}_{B}=\sum_{i=j}^{n_{max}}|v_j\rangle_{B}\langle 
v_j|_{B}$, where $n_{max}$ refers to the limited amount of 
guided modes present in the specific three-slab waveguide 
considered. For fixed values of $\gamma_0$ and 
$\sigma_{0}$, the spatial profile of photon B is the same, but 
the spatial profiles of the guided modes $\{v_{j}\}$ differ in 
their sizes for waveguides with different core size. Modes of the 
Schmidt decomposition $\{g_{j}(y)\}$ and guide modes in the 
TSW $\{v_{j}(y)\}$ can be ordered by its mode order 
($j=1,2,...$), with modes with the same order having similar
spatial shapes. In order to maximize the spatial overlap between
the Schmidt modes and the guided modes, we include an 
imaging system designed to maximize the overall spatial overlap
factor $F=\prod_j |d_{jj}|$. Fig.~\ref{Fig:3}(b),(c) and (d) show
the overall spatial overlap factor as a function of the
magnification factor (Z) of the imaging system for five different
three-slab waveguides which support $1$, $3$, $5$, $10$, and 
$15$ guided modes, respectively.  For instance, for 
$\sigma_0=1 \mu$m and $\gamma_0=3$, the optimum 
magnification factors are $0.55$, $0.82$, $1.13$, $1.61$, and 
$2.03$.

In contrast, since  we are interested in the Anderson
localization of photon A after propagation in the disordered
waveguide array, the spatial profile of photon A is detected by an
intensified coupled-charge detector (iCCD), which allows us to 
detect electromagnetic signals at the single-photon level. 
Detection of a photon in each pixel of the iCCD is represented via 
the photon-number operator $
\hat{n}_A\left(x\right)=\hat{E}^{(-)}_A\left(x\right)\hat{E}_A^{(+)}
\left(x\right)$. After detection of photon $B$, the spatial shape
of the photon A at distance $z$ in the WGA is described by the
photon-number spatial distribution
\begin{eqnarray}
& & p_{A}(x) = Tr_A \left[\hat{\rho}_A
\hat{n}_{A}\left(x\right)\right]=
\sum_{m,n}\sqrt{\lambda_{m}\lambda_{n}} I\left( m,n \right)
 \nonumber \\
& & \times\sum_{i,j} c_{mi} c^{*}_{nj} \exp \left\{ iz\left(
\kappa_{i}-\kappa_{j}
\right)\right\} u_{i}\left(x\right)u^{*}_{j}\left(x\right),
\label{Eq:12}
\end{eqnarray}
where $I \left( m,n \right)=\sum_{j} d_{mj} d^{*}_{nj}$. The
width of photon $A$ can be characterized by its effective beam
width
\begin{equation}
w_{eff}=\left\langle \frac{\left[\int dx p_{A}
(x)\right]^2}{\int
dx
p_{A}^2(x)}\right\rangle, \label{effective_width}
\label{Eq:13}
\end{equation}
where $\langle\rangle$ refers to averaging over an ensemble of
random realizations of a disordered WGA.

In order to analyze the results presented in Sec. III, it is
important to take into account that the beam size $\sigma_0$ and
the incoherence $\gamma_0$ of photon $A$, defined in Eqs.
(\ref{Eq:4}) and (\ref{Eq:5}), corresponds to values before
projection and detection of photon $B$. Therefore, after filtering
mediated by the spatial mode projection of photon B using the 
TSW, the first-order correlation function of  photon A at the input 
of WGA is written
\begin{equation}
G_A^{(1)}(x,x')=\sum_{m,n}\sqrt{\lambda_n\lambda_m}I\left(m,n\right)f_n\left(x\right)f^{*}_m\left(x'\right).
\label{Eq:14}
\end{equation}
One can obtain the values of $\sigma$ and $\gamma$ for photon $A$
via equations Eq.~(\ref{Eq:17}), Eq.~(\ref{Eq:19}) and
Eq.~(\ref{Eq:20}) in the the Appendix.

\begin{figure}[t!]
\centering
\includegraphics[width=8.70cm]{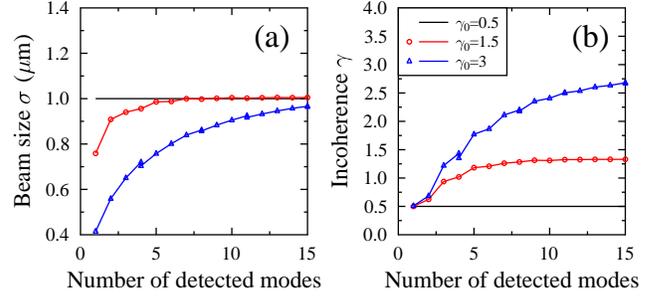}
\caption{(Color online) (a) Beam size $\sigma$ and (b) 
incoherence $\gamma$ of photon A when photon B propagates 
in  different TSWs, and afterwards is detected. Each TSW is 
designed to support a different number of guided modes, as 
indicated in the axis. We consider three different two-photon 
states (see legend in (b)), characterized by $\gamma_0=0.5$, 
1, and 3. In all cases, $\sigma_0\,=\,1\,\mu m$. }
 \label{Fig:4}
\end{figure}

If photon B propagates in a TSW that supports a single 
propagating mode, the size of photon A will corresponds to the 
size of that single mode, independently of the value of 
$\sigma_0$. When other 
modes are added via an increase of the guiding capability of TSW,
the beam size $\sigma$ reaches its initial value $\sigma_0$, as it
is shown in Fig.~\ref{Fig:4}(a) for a photon with $\sigma_0=1
\mu$m. A similar behavior of the value of $\gamma$ is also 
shown in Fig.~\ref{Fig:4}(b), where a strong dependence on the
effectiveness of the coupling to the TSW is observed. When
coupling to a single mode, $\gamma=0.5$, independently of the
value of $\gamma_0$. When the number of propagating modes 
in TSW is enlarged, the value of $\gamma$, even though it is 
smaller than $\gamma_0$, also converges to $\gamma_0$, 
since now propagation in the waveguide does not effectively 
filter the input state.

\begin{figure}[t]
\centering
\includegraphics[width=8.70cm]{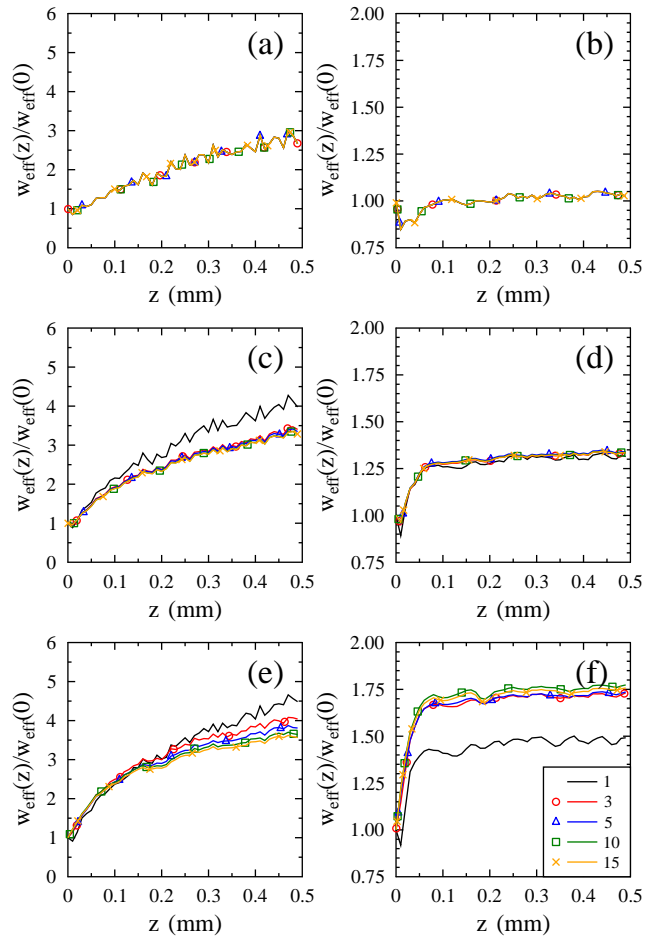}
\caption{(Color online) Spreading of the size of photon A after
propagation in the WGA, as given by the ratio
$w_{eff}(z)/w_{eff}(0)$. Photons A and B are part of a
two-photon state with three different values of the amount of
entanglement, but with the same value of $\sigma_0$ = 1 $\mu
m$. (a) and (b) correspond to a non-entangled two-photon state
with $\gamma_0=0.5$ (K=1). (c) and (d) corresponds to an
entangled state with $\gamma_0=1.5$ (K=3), while for (e) and
(f) we have $\gamma_0=3$ (K=6). (a),(c), and (e) correspond
to the propagation of photon A in a non-disordered WGA,
while (b),(d), and (f) corresponds to the propagation of photon A
in a disordered WGA with $\sigma=0.02$. We present averaged
results obtained over 100 different realizations of WGA. The
curves in all plots represent propagation of photon B in different
TSW which support distinct amounts of guided modes, as
shown in the legend in (f). This legend is valid for all plots.}
\label{Fig:5}
\end{figure}

\section{Results}
For the sake of comparison, we first consider a separable
two-photon state ($K=1$), so the Schmidt decomposition contains a
single mode, as shown in Fig.~\ref{Fig:1}(a). Photon A is in a
first-order coherent state, and since there is no entanglement,
there is also no dependence on the characteristics of the
propagation of photon A on photon B being projected and 
detected. As expected, when no disorder is considered 
($\delta=0$), photon A diffracts the least in comparison to other 
cases considered in Figs.\ref{Fig:5}(c) and (e), which 
corresponds to entangled paired photons.  When disorder is 
introduced ($\delta=0.02$), photon A turns out to be localized, 
with the size of the output probability distribution being almost 
equal to the input probability. Anderson localization is the result 
of the coupling of photon A to localized guided modes of the 
disordered WGA $\{u_{i}(x)\}$.

We now consider two examples with two-photon entangled 
states with  $\gamma_0=1.5$ and 3. This corresponds to 
two-photon states with Schmidt number $K=3$ and 6, and 
entropy of entanglement $E=2$ and 3.021. The Schmidt 
decompositions are shown in Fig.~\ref{Fig:1}(b) and (c). Unlike 
the coherent case ($\gamma_0$=0.5), the size of photon A 
depends on the amount of propagating modes of the TSW used. 
This phenomenon is more visible with the ordered WGA, as 
shown in Fig.~\ref{Fig:5}(c) and (e). Note that each Hermite 
function $\{f_i\}$ for $i>1$ contains high spatial components 
that spread even faster than the narrow Gaussian profile given 
by $f_1$, but in the overall, they might have a smaller impact on 
the final size of photon A due to decreasing weights 
$\lambda_j$ for a given state.

For a disordered WGA with $\delta=0.02$  the effect of the
partially coherent nature of photon A values on its propagation is 
more visible, as seen in Fig.~\ref{Fig:5}(d) and (f). The lower
the degree of coherence, the broader is the output effective 
width of the spatially localized photon A. Moreover, Hermite 
functions $\{f_i\}$ with increasing order localize with a higher 
ratio $w_{eff}(L)/w_{eff}(0)$ than the fundamental Hermite 
function $f_1$. Our calculations also predict a noticeable 
dependence of the amount of localization expected, shown in 
Fig.~\ref{Fig:5}, on important experimental values such as the 
magnification factor of the imaging system or the effectiveness 
of the coupling to the TSW. Therefore, if one were to use a 
different optimization function $F$ for the imaging system, 
differences in Fig.~\ref{Fig:5}(d) and \ref{Fig:5}(f) would be 
more visible.

\section{Conclusion}
We have presented an experimental scheme for the observation
of transverse Anderson localization of partially coherent light with
a tunable degree of coherence. The degree of coherence is
tuned by injecting one photon of a fully coherent two-photon
entangled state in a waveguide with a finite and controllable
amount of propagating modes. The system can be integrated on
a semiconductor chip, since both the disordered waveguide array
(WGA) and the three-slab waveguide (TSW) considered were
designed with this goal in mind. Therefore our proposal is
experimentally feasible taking into an account current mature
semiconductor technologies.

\section*{Acknowledgement}
J.S. thanks R. de J. Le\'{o}n-Montiel for useful discussions. We
acknowledge support from the Spanish government projects
FIS2010-14831 and Severo Ochoa, and from Fundaci\'o Privada
Cellex, Barcelona. J.S. acknowledges projects PrF-2013-006 of
IGA UP Olomouc, CZ.1.07/2.3.00/30.0004 and
CZ.1.07/2.3.00/20.0058 of M\v{S}MT \v{C}R. J.P. thanks
projects CZ.1.05/2.1.00/03.0058 and
CZ.1.07/2.3.00/20.0058 of M\v{S}MT \v{C}R.

\section*{Appendix}

\subsection{Quantifying the first-order coherence of the single photon}
The two-photon states given by Eqs. (\ref{two-photon-state1}) 
and (\ref{two-photon-state2}) describes a generally entangled 
state.  The density matrix that characterizes the quantum state 
of one of the photons that constitute the pair, for instance 
$\hat{\rho}_A$ for photon A, is obtained by tracing out the 
variables describing photon B, so
\begin{equation}
\hat{\rho}_{A}=\int dx\int dx' \rho_{A}(x,x')|x\rangle_A\langle
x'|_A, \label{Eq:15}
\end{equation}
where
\begin{equation} \rho_A(x,x')=\int dy
\Psi\left(x,y\right)\Psi^{*}\left(x',y\right) . \label{rho_A}
\end{equation}
Notice that $\rho_A(x,x')$ is the well-known first-order
correlation function $G_A^{(1)}(x,x')$ of photon A, defined as
\begin{equation}
G_A^{(1)}(x,x')={\rm Tr}\left[\hat{\rho}_A
\hat{E}_A^{(-)}\left(x\right)
\hat{E}_A^{(+)}\left(x'\right)\right],
\end{equation}
where $ \hat{E}_A^{(+)}$ and $ \hat{E}_A^{(-)}$ are the positive-
and negative-frequency electric-field operators
\cite{mandel_book}. The first-order correlation function for
photon B is defined similarly.

Making use of Eqs. (\ref{gaussian_mode_function}) and
(\ref{rho_A}), we obtain
\begin{eqnarray}
G_A^{(1)}(x,x')& \sim
&\exp\left[-\left(\alpha+\beta\right)x^{2}-\left(\alpha+\beta\right)x'^{2}\right.\nonumber\\
&
&\left.+\frac{\left(\alpha-\beta\right)^{2}}{2\left(\alpha+\beta\right)}\left(x+x'\right)^{2}\right].
\label{Eq:16}
\end{eqnarray}
The Gaussian form of the two-photon amplitude, as defined in
Eq.~(\ref{Eq:3}), allows us to quantify the width of photon A in
the position space using $G_A^{(1)}(x,x')$. The rms spatial width
of photon A is
\begin{equation}
\label{sigma0} \sigma^{2}=\frac{\int dx\,
x^{2}G_A^{(1)}(x,x)}{\int dx\,
G_A^{(1)}(x,x)}=\frac{\alpha+\beta}{16\alpha\beta}.
\label{Eq:17}
\end{equation}

The two-photon amplitude $\Psi(p,q)$ in the transverse wave-number
domain is equal to
\begin{equation}
\Phi(q,k) \sim
\exp\left[-\frac{\left(q+k\right)}{16\alpha}^{2}-\frac{\left(q-k\right)^{2}}{16\beta}\right].
\label{Eq:18}
\end{equation}
Similarly to the case considered above, the first-order
correlation function in the transverse wave-number domain reads
\begin{equation}
G_A^{(1)}(q,q')= {\rm
Tr}\left[\hat{\rho}_A\,\hat{a}_A^{\dagger}(q)\hat{a}_A(q')\right].
\end{equation}
One can calculate the rms width of photon A in the transverse
wave-number domain as
\begin{equation}
W^{2}=\frac{\int dq\, q^{2}G_A^{(1)}(q,q)}{\int dq\,
G_A^{(1)}(q,q)}=\alpha+\beta.
\label{Eq:19}
\end{equation}
Here we quantify the first-order coherence of photon A
as the product of its spatial beam width ($\sigma$) by its width
in the transverse wavevector domain ($W$)
\begin{equation}
\gamma =\sigma W=\frac{\alpha+\beta}{4\sqrt{\alpha\beta}},
\label{Eq:20} \label{gamma}
\end{equation}
this parameter $\gamma$ represents the amount of 
incoherence. For more details concerning quantification of 
coherence, see \cite{Perina1991}.  Making use of Eqs. 
(\ref{Eq:17}) and (\ref{Eq:20}) one easily obtains Eqs. 
(\ref{Eq:4}) and (\ref{Eq:5}) in the main text. The minimum 
value of $\gamma$ is $\gamma=0.5$. It corresponds to a 
separable two-photon state with $\alpha=\beta$. In this case, 
photon A (and photon B) shows first-order coherence. For 
entangled states, photon A is described by an incoherent 
superposition of Hermite-Gauss modes, whose number 
increases with a corresponding increase of the degree of
entanglement between photons A and B. Therefore, increasing 
values of $\gamma$ correspond to photons with a lower degree 
of coherence.


\begin{thebibliography}{99}

\bibitem{Anderson1958Absence} P. W. Anderson, Phys. Rev.
\textbf{109}, 1492 (1958). %1

\bibitem{Anderson2009PhysToday} A. Lagendijk, B. van
Tiggelen, and D. S. Wiersma, Phys. Today \textbf{62}, 24
(2009). %2

\bibitem{Cutler1969Observation} M. Cutler and N. F. Mott,
Phys. Rev. \textbf{181}, 1336 (1969). %3

\bibitem{Chabe2008Experimental} J. Chab\'e,  G. Lemari\'e, B.
Gr\'emaud, D. Delande, P. Szriftgiser, and J. C. Garreau, Phys.
Rev. Lett. \textbf{101}, 255702 (2008). %4

\bibitem{Billy2008Direct} J. Billy, V. Josse, Z. Zuo, A. Bernard, B.
Hambrecht, P. Lugan, D. Cl{\'e}ment, L. Sanchez-Palencia, P.
Bouyer, and A. Aspect, Nature \textbf{453}, 891 (2008). %5

\bibitem{Roati2008Anderson} G. Roati, C. D\`{ }Errico, L.
Fallani, M. Fattori, C. Fort, M. Zaccanti, G. Modugno, M.
Modugno, and M. Inguscio, Nature \textbf{453}, 895 (2008).
%6

\bibitem{Hu2008Localization} H. Hu, A. Strybulevych, J. H.
Page, S. E. Skipetrov, and B. A. van Tiggelen, Nat. Phys.
\textbf{4}, 945 (2008). %7

\bibitem{Raedt1989Transverse} H. De Raedt, A. Lagendijk, and
P. de Vries, Phys. Rev. Lett. \textbf{62}, 47 (1989). %8

\bibitem{Wiersma1997Localization}  D. S. Wiersma, P. Bartolini,
A. Lagendijk, and R. Righini, Nature \textbf{390}, 671 (1997).
%9

\bibitem{Schwartz2007Transport} T. Schwartz, G. Bartal, S.
Fishman and M. Segev, Nature \textbf{446}, 52 (2007). %10

\bibitem{Schreiber2011Decoherence} A. Schreiber, K. N.
Cassemiro, V.  Poto\ifmmode \check{c}\else \v{c}\fi{}ek, A.
G\'abris, I. Jex, and Ch. Silberhorn, Phys. Rev. Lett.
\textbf{106}, 180403 (2011). %11

\bibitem{Crespi2013Anderson} A. Crespi, R. Osellame, R.
Ramponi, V. Giovannetti, R. Fazio, L. Sansoni, F. De Nicola, F.
Sciarrino, and P. Mataloni, Nat. Photon. \textbf{7}, 322 (2013).
%12

\bibitem{Lahini2008Anderson}Y. Lahini, A. Avidan, F. Pozzi, M.
Sorel, R. Morandotti, D. N. Christodoulides, and Y. Silberberg,
Phys. Rev. Lett. \textbf{100}, 013906 (2008). %13

\bibitem{Aharonov1993Quantum} Y. Aharonov, L. Davidovich,
and N. Zagury,  Phys. Rev. A \textbf{48}, 1687 (1993). %14

\bibitem{Svozilik2012Implementation} J. Svozil\'{i}k, R. de J.
Le\'{o}n-Montiel, and J. P. Torres, Phys. Rev. A \textbf{86},
052327 (2012). %15

\bibitem{Yin2008Quantum} Y. Yin, D. E. Katsanos, and S. N.
Evangelou, Phys. Rev. A \textbf{77}, 022302 (2008). %16

\bibitem{PerinaJr2009b} J. {Pe\v{r}ina~Jr.}, M. Centini, C.
Sibilia, and M. Bertolotti, J. Russ. Laser Res. \textbf{30}, 508
(2009). %17

\bibitem{PerinaJr2009c} J. {Pe\v{r}ina~Jr.}, M. Centini, C.
Sibilia, and M. Bertolotti, Phys. Rev. A \textbf{80}, 033844
(2009). %18

\bibitem{Inui2004Localization} N. Inui, Y. Konishi, and N. Konno,
Phys. Rev. A \textbf{69}, 052323 (2004). %19

\bibitem{Shor1994Algorithms} P. W. Shor,  in 
\textit{ Proc. 35nd Annual Symposium on Foundations of 
Computer Science}, Santa Fe, 1994, edited by S. Goldwasser 
(IEEE Computer Society Press, Washington, 1994), p. 124. %20

\bibitem{Grover1996Afast} L. K. Grover,  in \textit{Proc. 28th 
Annual ACM Symposium on Theory of  Computing}, Philadelphia, 
1996, edited by  G. L. Miller (ACM New York, 1996),  p. 212. 
%21

\bibitem{Capeta2011Anderson} D. {\v{C}}apeta, J. Radi{\'c},
A. Szameit, M. Segev, and H. Buljan, Phys. Rev. A \textbf{84},
011801 (2011). %22

\bibitem{mandel_book} L. Mandel and E. Wolf, {\em Optical
coherence and quantum optics}, Cambridge University Press,
New York, 1995. %23

\bibitem{segev2001} D. N. Christodoulides, E. D. Eugenieva, T.
H. Coskun, M. Segev, and M. Mitchell, Phys. Rev. E \textbf{63},
035601 (2001). %24

\bibitem{glauber1966} U. M. Titulaer and R. J. Glauber, Phys.
Rev. \textbf{145}, 1041 (1966). %25

\bibitem{victor2009} V. Torres-Company, A. Valencia, and J. P.
Torres, Opt. Lett. \textbf{34}, 1177 (2009). %26

\bibitem{Lahani2010Quantum}
Y. Lahini, Y. Bromberg, D. N. Christodoulides, and Y. Silberberg,
Phys. Rev. Lett. \textbf{105}, 163905 (2010). %27

\bibitem{Abouraddy2012Anderson} A. F. Abouraddy, G. Di
Giuseppe, D. N. Christodoulides, and B. E. A. Saleh, Phys. Rev. A
\textbf{86}, 040302 (2012). %28

\bibitem{DiGiuseppe2013Einstein} G. Di Giuseppe, L. Martin, A.
Perez-Leija, R. Keil, F. Dreisow, S. Nolte,  A. Szameit, A. F.
Abouraddy, D. N. Christodoulides, and B. E. A. Saleh, Phys. Rev.
Lett. \textbf{110}, 150503 (2013). %29

\bibitem{progress2011} J. P. Torres, K. Banaszek and I. A. Walmsley,
Progress in Optics \textbf{56} (chapter V), 227 (2011).
% Engineering Nonlinear Optic Sources of Photonic Entanglement

\bibitem{chan_eberly} K. W. Chan and J. H. Eberly,  arXiv
preprint quant-ph/0404093v2 (2004). %30

\bibitem{PerinaJr2008} J. Pe\v{r}ina~Jr., Phys. Rev. A
\textbf{77}, 013803 (2008). %31

\bibitem{Gehrsitz2000The} S. Gehrsitz and F. K. Reinhart and
C. Gourgon and N. Herres and A. Vonlanthan and H. Sigg,
J. Appl. Phys. \textbf{87}, 7825 (2000). %32

\bibitem{Karbasi2013Modal} S. Karbasi, K. W. Koch, A. Mafi, J.
Opt. Soc. Am. B \textbf{30}, 1452 (2013). %33

\bibitem{Jin2002Finite} Jianming Jin, {\em The finite element
method in electromagnetics}, John Wiley and sons, New York,
2002. % 34

\bibitem{Snyder1983Optical} A. W. Snyder and J. Love, {\em
Optical Waveguide Theory}, Springer, 1983. %35

\bibitem{Perina1991} J. Pe\v{r}ina, {\em Quantum Statistics of Linear and Nonlinear Optical
Phenomena} (Kluwer, Dordrecht, 1991).

\end{thebibliography}
\end{document}